\documentclass[%
 reprint,
 amsmath,amssymb,
 aps,
prb,
]{revtex4-1}

\usepackage{graphicx}
\usepackage{dcolumn}
\usepackage{bm}

\usepackage{color}
\definecolor{del}{RGB}{255,20,147}
\definecolor{ins}{RGB}{0,139,0}

\begin{document}

\preprint{APS/123-QED}

\title{Room Temperature Coherent Spin Alignment of Silicon Vacancies in 4$H$- and 6$H$-SiC}

\author{Victor A. Soltamov}
\author{Alexandra A. Soltamova}
\author{Pavel G. Baranov}
\email{Pavel.Baranov@mail.ioffe.ru}
\affiliation{%
Ioffe Physical-Technical Institute, St. Petersburg, 194021 Russia
}%

\author{Ivan I. Proskuryakov}
\affiliation{%
Institute of Basic Biological Problems RAS, Pushchino 142290, Russia
}%

\date{\today}%

\begin{abstract}

We report the realization of the optically induced inverse population of the ground-state spin sublevels of the silicon vacancies ($V_{\mathrm{Si}}$) in silicon carbide (SiC) at room temperature. The data show that the probed silicon vacancy spin ensemble can be prepared in a coherent superposition of the spin states. Rabi nutations persist for more than 80~$\mu$s. Two opposite schemes of the optical alignment of the populations between the ground-state spin sublevels of the silicon vacancy upon illumination with unpolarized light are realized in 4H- and 6H-SiC at room temperature. These altogether make the silicon vacancy in SiC a very favorable defect for spintronics, quantum information processing, and magnetometry.

\begin{description}
\item[PACS numbers]
61.72.Hh, 71.55.-i, 76.70.Hb, 61.72.jd
\end{description}
\end{abstract}

\pacs{Valid PACS appear here}
\maketitle

Detection and manipulation of the spin states in solids at room temperature is a basis of the emerging fields of quantum information processing and spintronics. The first system on which such manipulations were realized at room temperature was the nitrogen-vacancy (NV) center in diamond. Owing to its unique optical excitation cycle that leads to the optical alignment of triplet sublevels of the defect ground state, single NV center can be easily initialized, manipulated and readout by means of optically detected magnetic resonance (ODMR) \cite{Gruber_Sci_97_1,Jelezko_PSS_06_2,Jel_Popa_APL_02_3,Niz_Jel_PhB_03_4}. A search for systems possessing unique quantum properties of the NV defect in diamond that can extend the functionality of such systems seems to be a very promising objective. 

On the basis of theoretical predictions and experimental data several other centers were proposed as candidates comparable with the NV center in diamond. Among them such centers as the nitrogen-vacancy center \cite{Weber_PNAS_2010_6,DuVincenzo_Nature_2010_7}, silicon-carbon divacancy \cite{Baranov_JETP_05_8,Koehl_Nature_11_9} and silicon vacancy ($V_{\mathrm{Si}}$) \cite{Baranov_JETP_07_10,Baranov PRB_11_11} in silicon carbide were proposed. Recently it was experimentally shown that several defect spin states in 4H-SiC can be optically addressed and coherently controlled at temperatures from 20 to 300 K \cite{Koehl_Nature_11_9}.

Silicon carbide is a wide-band-gap semiconductor with a well-developed growth and doping technology that opens wide possibilities for scalable applications. The isotopic engineering of SiC crystals can be performed through the sublimation crystal growth\cite{Baranov_isotope2_13} in order to reduce the abundance of $^{29}$Si and $^{13}$C isotopes having nonzero nuclear spins.  SiC can be crystallized in many different polytypes that arise from differences in the stacking sequence of the Si and C layers. The most common polytypes are 4H- and 6H–SiC. In 4H-SiC, two nonequivalent crystallographic positions exist: one hexagonal and one quasicubic called $\textit{h}$ and $\textit{k}$, respectively. In 6H-SiC, three nonequivalent positions are formed: one hexagonal ($\textit{h}$) and two quasicubic ones ($\textit{k}$1 and $\textit{k}$2) [see Figs. 1(a) and 3(a)].  Because of the difference in the surrounding environments, a defect located at the $h$ and $k$ sites often has different properties.

Vacancies are the primary defects in SiC incorporated at various sites in different polytypes. Both photoluminescence and electron paramagnetic resonance (EPR) spectra of these centers vary  depending on their position in the crystal lattice. Unusual polarization properties of various vacancy defects in SiC were observed by means of EPR under optical excitation and reported for the first time in the work of Vainer and Il'in \cite{Vainer_FTT_81_5}. Observation of the low temperature optical spin alignment of the $V_{\mathrm{Si}}$ ground state and zero-field ODMR studies at 1.4~ K were reported in Ref. [\cite{Baranov PRB_11_11}]. In this study we show that two opposite schemes for the optical alignment of the spin sublevels in the ground state of silicon vacancies in 4H- and 6H-SiC can be realized even at room temperatures and that a spin ensemble can be prepared in a coherent superposition of the spin states.  

\begin{figure}
\includegraphics{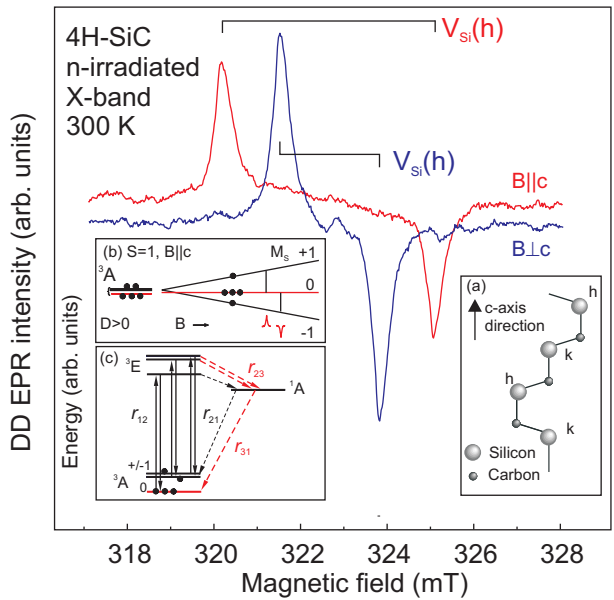}
\caption{\label{Fig1_rev} cw DD-EPR spectra recorded at room temperature in the 4H-SiC crystal under excitation at the wavelength of 890 nm for B$\|\textit{c}$ and B$\perp$$c$ orientations. (a) Structure of the 4H-SiC polytype. (b) Zeeman level diagram for the $V_{\mathrm{Si}}$ ground state ($S$=1) with the $M{_S}$~=~0 sublevel predominantly populated due to the optical alignment. (c) Seven-level model interpreting the optical alignment of the ground-state sublevels in zero magnetic field. 
}
\end{figure}

Crystals of two main SiC polytypes were studied in the present work: 4H-SiC and 6H-SiC. Silicon vacancies were introduced by irradiation with fast neutrons at room temperature with a dose of 10$^{15}$~cm$^{-2}$~-~10$^{16}$~cm$^{-2}$. The samples, in the shape of platelets, had dimensions of about 3$\times$4$\times$0.4~mm$^{3}$ and were oriented for rotation in the $\{$11\={2}0$\}$ plane. The concentration of vacancies under the study in both samples was $\sim$10$^{15}$~cm$^{-3}$. The precise concentrations are not required for the following analysis.

The X-band ($\sim$9 GHz) direct-detection (DD) EPR technique \cite{Borovykh_JPCh_00_14} was used in our experiments. In DD-EPR the signals induced by optical flash are measured in direct absorption and emission mode in the continuous wave regime. The sample was selectively excited at 890 nm with 6 ns flashes (ca. 1.5 mJ per flash) from a parametric oscillator LP603 pumped by a Nd-YAG laser LQ 529B (Solar Laser Systems, Byelorussia), which was operated at a repetition rate of 11 Hz and 0.8 nm FWHM. The EPR signal obtained after the microwave mixer was amplified by wide-band amplifiers and sampled by a boxcar integrator (SR 250, Stanford Research Systems) triggered by the flashes. The absolute microwave power value was estimated as ca. 5 mW at 10 dB attenuation. The 50 ns time resolution allowed signal detection shortly after the excitation laser flash, before any significant relaxation between spin sublevels had taken place. To increase the signal-to-noise ratio, up to 50 field scans were averaged.

cw DD-EPR signals of the $V_{\mathrm{Si}}$ recorded at room temperature in the 4H-SiC crystal for two orientations of the magnetic field- parallel (B$\|\textit{c}$) and perpendicular (B$\perp$$c$) to the $\textit{c}$ axis- are shown in Fig.1. The sample was selectively excited by optical flash into the absorption band of the $V_{\mathrm{Si}}$ at 890~nm. The same type of EPR signal was also observed with the excitation light at wavelengths of 865, 900 and 917~nm. Observed lines with the zero-field splitting parameter $D$ of 22$\times$10$^{-4}$~cm$^{-1}$ (65.9 MHz) labeled as $V_{\mathrm{Si}}$ belong to the silicon vacancy in the $h$ site of the SiC lattice (Fig.1(a)). The lines are accompanied by the pair of the hyperfine lines with a splitting of 0.29 mT due to the hyperfine interaction with 12 equivalent Si atoms of the second shell of the Si vacancy. The spin state of the $V_{\mathrm{Si}}$ giving rise to this splitting is still under debate. Two models were proposed: (i) neutral silicon vacancy with $S$=1 and (ii) low-symmetry modification of the well-studied negatively charged Si vacancy in the regular environment with $S$=3/2\cite{Mizuochi_PRB_0205_19}. For the purpose of further discussion we will assume that these signals belong to the $V_{\mathrm{Si}}$ with $S$=1 and discuss the case of $S$=3/2 at the end. 

Observed DD-EPR spectra can be described with standard spin Hamiltonian
\begin{equation}
    \hat{H}=\textsl{g}\mu_{B}\emph{$\textsl{B}$}\emph{$\textsl{$\hat{S}$}$}+D(\hat{S}_{z}^{2}+1/3S(S+1))
     \label{pauli}
\end{equation}
where $\mu_{B}$ is the Bohr magneton, $\textsl{g}$ is the isotropic $\textsl{g}$ factor ($\textsl{g}$=2.0032), and $S$=1. For 4H-SiC the zero-field splitting parameter $D$ of the $V_{\mathrm{Si}}$ was found to be 22$\times$10$^{-4}$ cm$^{-1}$ (65.9 MHz) for the $\textit{h}$ site \cite{Bardel_Batt_PRB_00_17}. 

For the high-field transitions, emission instead of absorption is detected; thus, we can conclude that optical excitation at room temperature results in the establishment of the inverse population between certain spin sublevels, i.e. optical alignment of the ground spin sublevels. The Zeeman levels diagram for the $V_{\mathrm{Si}}$ ground state is shown in Fig. 1(b). The populations of the ground-state energy sublevels under optical pumping are indicated by different numbers of filled circles. The emission observed for the high-field component of the EPR spectra is due to predominant population of the $M{_S}$=0 sublevel.

To explain the photokinetic process leading to the spin alignment under optical pumping, we adopt the seven-level model proposed for the NV centers in diamond\cite{Manson_PRB_06_18}, which suggests the existence of the nonradiative recombination channel. The level diagram shown in Fig.~1(c) comprises a ${^ 3}$A ground state, a ${^ 3}$E excited state for which only three lower sublevels are depicted, and one singlet ${^ 1}$A state. Optical transition between the ${^ 3}$A and ${^ 3}$E states is spin conserving (solid lines in Fig.~1(c)). In addition, triplet-singlet intersystem crossing due to the spin-orbit coupling between ${^ 3}$E and ${^ 1}$A levels occurs. The centers in the $M{_S}$=$\pm$1 sublevels have significantly higher probability to undergo the ISC; thus, the rate $r{_{23}}$ of the nonradiative transition from the $M{_S}$=$\pm$1 sublevels of the excited ${^3}$E state to the metastable singlet ${^1}$A state is much larger than the rates of the transition from the $M{_S}$=0 sublevel. Subsequent decay from the ${^1}$A to the $M{_S}$=0 sublevel of the ${^3}$A ground state also occurs with a higher rate ($r{_{31}}$) than that between ${^1}$A and $M{_S}$=$\pm$1. As a result the $M{_S}$=0 sublevel becomes predominantly populated. As the $M{_S}$=0 sublevel has a higher probability of fluorescence due to the nonradiative nature of the intersystem crossing, the fluorescence intensity between ${^ 3}$E and ${^ 3}$A is spin-dependent. The discussed scheme is true for the $V_{\mathrm{Si}}$ center in the $h$ site  of the 4H-SiC lattice at room temperature and is similar to the NV defects in diamond. This type of polarization should result in a giant decrease of the photoluminescence intensity in ODMR experiments in zero magnetic field.

\begin{figure}
\includegraphics{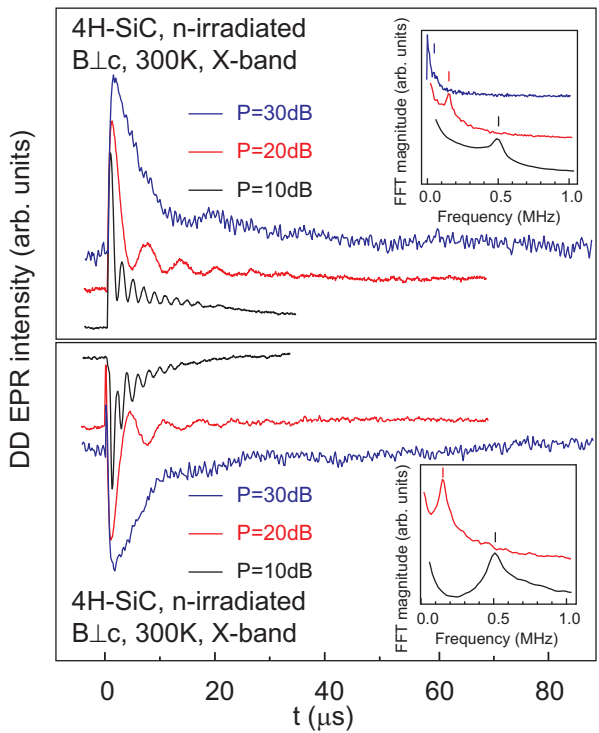}
\caption{\label{fig2}Transient nutations for the $V_{\mathrm{Si}}$ center in 4H-SiC at room temperature are shown for three values of microwave power: The positive traces stand for absorption ($B_0$=321.5~mT), and the negative traces for emission ($B_0$=323.8~mT) of microwaves. Absolute mw power value is estimated as ca. 5~mW at 10 dB attenuation. (Inserts) Corresponding fast Fourier transform (FFT).
}
\end{figure}

In Fig. 2, the transient nutations for the $V_{\mathrm{Si}}$ center in 4H-SiC at room temperature are shown for three values of microwave power $P$ at $B_0$=321.5~mT (top) and $B_0$=323.8~mT (bottom). In the experiment, the microwave absorption intensity resonant with the spin transitions for two EPR lines [Fig. 1, (B$\perp$$c$)] is monitored as a function of the time duration (t) after optical flash. 
The transition nutation decays due to inhomogeneity of the \textbf{\emph{$B_{1}$}} microwave field over the sample. In addition, the resonance frequencies are also spread around some mean value of resonance Larmor frequency $<$$\omega{_0}$$>$ leading to the inhomogeneous line broadening.

Clearly, the observed oscillatory behavior demonstrates that the probed $V_{\mathrm{Si}}$ spin ensemble can be prepared in a coherent superposition of the spin states at resonant magnetic fields at room temperatures. The population difference of spin states becomes modulated in time with the Rabi frequency given by $\omega{_1}$=~$\gamma{_e}$B${_1}$, where $\gamma{_e}$ is the gyromagnetic ratio for the electron. The Fourier transforms corresponding to observed oscillations are presented in the insets in Fig. 2. The Rabi frequencies are $\omega{_1}$=0.02, 0.16, and 0.5 MHz at $P$=30, 20, and 10 dB, respectively. Rabi oscillations decay with a characteristic time constant $\tau_R$ that depends on the microwave power (Fig. 2).  Empirically, $\tau_R$ is generally smaller than the spin-lattice relaxation time ($T_1$), thus the lower limit of $T_1$ is about 80~$\mu$s at room temperature.

Another important factor that should be considered in order to determine the applicability of a system for quantum information processing is the degree of spin polarization. The degree of optical polarization induced by optical pumping in our experiments can be roughly estimated from the kinetics of the DD-EPR signal measured at high microwave power (Fig. 2). To eliminate possible light-induced artifact signals, the procedure of subtraction of high-field and low-field kinetics was performed. The initial intensity of the DD-EPR signal corresponds to the nonequilibrium population differences between ground spin state sublevels created by the optical flash. Microwave-induced relaxation of nonequilibrium population leads to the equilibrium spin sublevel population determined by Boltzmann statistics. From the ratio between the initial DD-EPR signal intensity at 10 dB and the signal at $\sim$30$\mu$s the value of spin polarization is estimated as $\sim$80\%.

Figure 3 shows cw DD-EPR signals in the 6H-SiC crystal detected at room temperature and 100 K in B$\perp\textit{c}$ orientation of the magnetic field. The sample is selectively excited by optical flash into the absorption band of the $V_{\mathrm{Si}}$. Vertical bars indicate the positions of the lines of the $V_{\mathrm{Si}}$ vacancy in the hexagonal $\textit{h}$ and two quasicubic sites ($\textit{k}$1 and $\textit{k}$2). The spectra are described with the standard spin Hamiltonian [Eq. (1)] with $D$~=~42.8$\times$10$^{-4}$~cm$^{-1}$ (128.3 MHz) for $\textit{h}$ and 9$\times$10$^{-4}$~cm$^{-1}$ (26.9 MHz) for $\textit{k}$ sites of 6H-SiC, respectively \cite{Bardel_Batt_PRB_00_17}. 

As can be seen from Fig.~3, another type of optical alignment is realized for the $V_{\mathrm{Si}}$ in the $h$ and $k$ sites in the 6H-SiC at room temperature\cite{Baranov_JETP_07_10,Baranov PRB_11_11}. To explain the emissive nature of the low-field transition, Zeeman levels diagram for the $V_{\mathrm{Si}}$ ground state is shown in the inset (Fig.~3). The different numbers of circles indicates the population differences between sublevels. Such population differences induced by the optical pumping can be explained if we suggest the existence of a spin-dependent nonradiative decay path via the ${^1}$E level in the optical cycle. The decay rate $r{_{23}}$ from the $M{_S}$=$\pm$1 sublevels of the excited ${^3}$E state to the metastable ${^1}$E state \cite{Manson_PRB_06_18} is again much larger than the rates of transition from the $M{_S}$=0 sublevel, but, in contrast to the $V_{\mathrm{Si}}$ in 4H-SiC, the rates of the transitions $r{_{31}}$ between the ${^1}$E state and $M{_S}$=0 sublevel of the ground state are much smaller than the rates of the transitions from the ${^1}$E to the $M{_S}$=$\pm$1 sublevels of the ground ${^3}$A state. Thus, the spin sublevels with $M{_S}$=$\pm$1 of the ${^3}$A ground state are predominantly populated. This type of polarization should result in a giant increase of the photoluminescence intensity in ODMR experiments in zero magnetic field.

\begin{figure}
\includegraphics{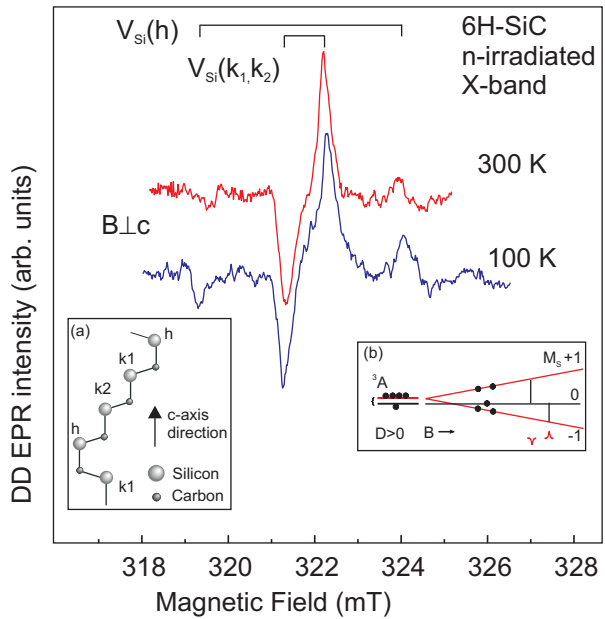}
\caption{\label{fig3}
cw DD-EPR spectra recorded at room temperature and 100~K in the 6H-SiC crystal under excitation by optical flash at 890 nm for B$\perp$$c$ orientation. (a) Structure of the 6H-SiC polytype; (b) Zeeman level diagram for the $V_{\mathrm{Si}}$ ground state ($S$=1); $M{_S}$~=~$\pm$1 sublevels predominantly populated due to the optical alignment.
}
\end{figure}

As was noted the spin state of the silicon vacancy is still under debate. However, similar alignment processes can be valid for the $S$=3/2 system. In principle the mechanism of spin alignment should remain almost the same with the only difference that the $M{_S}$=$\pm$1 and $M{_S}$=0 sublevels should be replaced by $M{_S}$=$\pm$3/2 and $M{_S}$=$\pm$1/2, respectively. For this model $M{_S}$=$\pm$1/2 sublevels should be equally populated and the existence of a metastable doublet state instead of a singlet state should be assumed.

In conclusion, optically induced alignment (polarization) of the ground-state spin sublevels of the Si vacancy in 4H- and 6H-SiC was observed for the first time at room temperature. In distinction from the known NV defect in diamond\cite{Gruber_Sci_97_1} and recently observed defects in SiC\cite{Koehl_Nature_11_9}, two opposite schemes for the optical spin alignment of $V_{\mathrm{Si}}$  in 4H- and 6H-SiC were realized at room temperature upon illumination with unpolarized light. The alignment schemes vary depending on the crystal polytype and crystallographic position of the silicon vacancy in the crystal lattice, as well as do their zero-phonon lines and zero-field splitting parameters. Thus, optical quantum computation and communication protocols for silicon vacancies located in the hexagonal and quasicubic sites of the SiC lattice should vary as well. Silicon vacancies are the primary defects in SiC. This center has an axial symmetry along the (0001) crystallographic direction, thus it is not necessary to control the defect orientation neither by application of transverse magnetic field, nor during the $V_{\mathrm{Si}}$ creation. Observed Rabi nutations persist for $\gtrsim$80~$\mu$s at room temperature and evidence that the probed $V_{\mathrm{Si}}$ spin ensemble can be prepared in a coherent superposition of the spin states at resonant magnetic fields at room temperature. Demonstrated spin properties of the silicon vacancies open up intriguing avenues for quantum computing and magnetic resonance imaging in the near-infrared optical and radio frequency bands.

This work has been supported by Ministry of Education and Science, Russia,
under Contracts No. 14.740.11.0048 and No. 16.513.12.3007, by the programs of
the Russian Academy of Sciences: "Spin Phenomena in Solid State
Nanostructures and Spintronics" and "Fundamentals of nanostructure and
nanomaterial technologies", and by the Russian Foundation for Basic
Research.

\end{document}